\documentclass[singlecolumn,pre,notitlepage,preprintnumbers,floatfix,superscriptaddress]{revtex4-1}
\usepackage{amssymb}
\usepackage{amsbsy}
\usepackage{amsmath}
\usepackage{bbm}
\usepackage{xcolor}

\usepackage{graphicx}
\usepackage{bm}
\usepackage{subfigmat}

\usepackage{amsmath}
\usepackage{verbatim}
\usepackage{color}
\usepackage{amssymb}
\usepackage{ulem}
\usepackage{soul}
\usepackage{natbib}
\usepackage{textcomp}

\begin{document}

\title{Flow induced intermittent transport shapes colloid filtration in complex media}

\author{Filippo Miele}
\affiliation{Institute of Earth Sciences, University of Lausanne, Lausanne 1015, Switzerland}

\author{Ankur Deep Bordoloi}
\affiliation{Institute of Earth Sciences, University of Lausanne, Lausanne 1015, Switzerland}

\author{Marco Dentz}
\affiliation{Spanish National Research Council (IDAEA-CSIC), Barcelona 08034, Spain}

\author{Herve Tabuteau}
\affiliation{University of Rennes, CNRS, IPR (Institut de Physique de Rennes)-UMR 6251, F-35000 Rennes, France}

\author{Ver\'{o}nica L. Morales}
\affiliation{Department of Civil and Environmental Engineering, University of California, Davis, USA}

\author{Pietro de Anna}
\affiliation{Institute of Earth Sciences, University of Lausanne, Lausanne 1015, Switzerland}
\email{pietro.deanna@unil.ch}
\email{filippo.miele@epfl.ch}
\email{aerials00@gmail.com}
\email{marco.dentz@csic.es}
\email{vermorales@ucdavis.edu}
\email{herve.tabuteau@univ-rennes.fr}

\maketitle

\section*{Abstract}

\noindent
The macroscopic phenomenon of filtration is the separation between suspended and liquid phases and it takes place in natural environments (e.g. groundwater, soil and hyporheic zone) and industrial systems (e.g. filtration plants, pharmaceutical industry and hospital care). Porous materials represent excellent filters since they are characterized by a large amount of solid surface to which flowing particles can attach and be retained. Colloidal filtration by porous media is governed by a complex interplay between transport dynamics through intricate pore structure and surface-mediated retention. Yet, classical approaches fail to capture key properties (such as the filter spatial heterogeneity) and experimental observations - e.g. non-exponential deposition profiles. A key limitation of such approaches lies in the assumption that the particle attachment to solid surfaces occurs at a constant rate over a given length scale, neglecting the intrinsic heterogeneity of the medium, i.e. pore size variability. Here, we develop a multiscale microfluidic model system to directly observe colloidal transport and deposition over more than three orders of magnitude in length - from tens of microns to a meter - within a porous architecture with controlled structural/physical heterogeneity. By tracking individual colloidal particles within the pores, we reveal an intermittent dynamics along each trajectory: particles alternate between long-range "flights" through pore channels and short-range localized "dives" near the grain surfaces. Within the dives attachment can occur at a constant rate, but the widely distributed flights size, during which attachment cannot occur, produces anomalous filtration, quantified by deposition profiles and breakthrough curves. The broad distribution of flights and dives spatial extension reflects the heterogeneity of the medium. We capture these properties of particle dynamics with a novel stochastic model based on a continuous time random walk (CTRW), constrained entirely by experimentally measured parameters. This model links the microscopic structure of the medium with macroscopic filtration, coupling complex fluid flow with colloidal surface interactions to more accurately describe non-exponential deposition profiles.

\section{Introduction}

\noindent
Colloid transport and retention is a crucial process in porous media that significantly impacts diverse environmental, biological, and industrial domains, encompassing abiotic and living entities. In terrestrial environments, for example, natural colloids that carry pollutants migrate through fractured rocks and soils, where they interact with minerals and microbial communities, influencing soil composition and contaminant dispersion \cite{bradford_transport_2013,grolimund_1996}. In the human body, viruses, microorganisms, and drug-carrying nanoparticles are transported through complex porous structures comprising the gut \cite{ejazi_mechanisms_2023,scheidweiler_spatial_2024}, extracellular matrices~\cite{malandrino_complex_2018}, and the vascular system \cite{blanco_principles_2015,corliss_methods_2019,jafarnejad_quantification_2019}. Furthermore, colloidal transport also plays a crucial role in filtration systems within many industrial and automotive contexts, where aerosol particles and pollutants dispersed in solvents are captured and retained inside or at the entrance of porous media \cite{tcharkhtchi_overview_2021,bradford2015determining}.\\ 

 \noindent
Colloid Filtration Theory (CFT) has traditionally provided a macroscopic description of colloidal transport within porous media using an advection-dispersion framework. CFT treats the underlying medium as homogeneous, and it considers a constant particle attachment rate over a fixed length scale characterizing the porous medium~\cite{Nelson2007,GinnLang2005}. For more than fifty years, studies have relied on this model to predict macroscopic transport behaviors, with complementary experiments, often conducted by injecting particles into a packed bed of spherical beads \cite{Happel1958}. While successful for relatively homogeneous media, CFT predictions have been challenged by observations in physically heterogeneous structures~\cite{Cortis1,Tuf_Elim2_2004, tufenkji2005breakdown} for which the structural medium properties cannot be described by a single length scale, but a wide distribution of values (e.g., variability in pore size, dead ends, fluid velocity distribution) that control enhanced colloid spreading and are not taken into account by CFT~\cite{redman_pathogen_2001}. Therefore, despite the ubiquity of this process, our understanding of the transport and retention mechanisms remains limited, highlighting the need of a new theoretical framework that incorporates the complex and heterogeneous features of porous media. \\

\noindent
Most of the relevant experimental studies have been conducted using columns packed with sand, soil, or rock samples to examine the effects of flow dynamics and realistic physical and chemical conditions on colloidal transport~\cite{johnson1996colloid,baygents_variation_1998,molnar2015predicting}. However, these configurations typically provide access only to macroscopic transport features, such as breakthrough curves (BTCs), representing particle flux through the medium, and Deposition Profiles (DPs), which quantify the spatial distribution of attached particles. Deposition profiles are usually obtained a-posteriori by sectioning the medium into thin slices, adding further experimental complexity~\cite{Li_EST2005}. Only a limited number of studies have captured pore scale processes of particle aggregation~\cite{perez_morphology_2020,gerberPRL2018} and temporal evolution of deposition profiles using confocal microscopy \cite{bizmark_multiscale_2020}, or X-ray Computed Tomography (CT)~\cite{patino_2023}. However, to discern the mechanisms that lead to said concentration signals, it is imperative to know the dynamics of a particle as it percolates the porous medium. Tracking particles trajectory in real time remains challenging due to the time-intensive nature of the scanning process and the spatial limitation introduced by the trajectory size compared to the filter size. Consequently, we lack a direct link between macroscopic particle filtration and microscopic particle transport across disordered porous media. \\

\noindent
The development of microfluidic techniques has allowed for the fabrication of transparent porous systems with controlled structure, allowing direct access to processes taking place under flow conditions~\cite{Bartolo_NatPhys2024, Dehkharghani2019Bacterial, deAnnaNaturePhys2021, waisbord_fluidic_2021}. This enables detailed studies of trajectory dynamics at the pore scale by tracking of individual particles across successive pores~\cite{scheidweiler_trait-specific_2020, waisbord_anomalous_2019}. Exploiting particle tracking, a few studies \cite{bordoloi_structure_2022,Scheidweiler2020} have attempted to connect macroscopic transport, such as BTCs, to the underlying pore-scale mechanisms, thereby elucidating the origins of anomalous particle transport (e.g., late time elution of the BTC and hyper-exponential or non-monotonic DP). While only a few studies to date have explored the impact of pore-scale hydrodynamics on Deposition Profiles across multiple scales~\cite{bizmark_multiscale_2020,pradel2024role}, the link between the detailed microscopic structure and macroscopic transport and filtration remains elusive. \\

 \noindent
Here, we develop a microfluidic setup, which allows to determine multi-scale transport and filtration of colloidal particles. On the one hand, the porous medium structure is heterogeneous with overall size $L = 1$~m and it is composed of cylindrical grains distributed in size ($L$ being about $15,000$ average pores $\overline{\xi}$). On the other hand, we resolve the local hydrodynamics using particle tracking experiments, showing that particle transport along trajectories has intermittent behavior jumping between two states, that we call flights and dives. The first corresponds to the state of a particle transported at the local fluid velocity, remaining at a sufficient distance from any solid surface to abstain from surface interaction. The second corresponds to a particle traveling close enough to a grain surface so that the colloid can interact with the surface and potentially attach. We demonstrate how the heterogeneity of the medium induces a broad distribution in the spatial extent of each state. We further show that this pore-scale retention mechanism controls macroscopic features of particle transport that are not captured by CFT (in terms of BTC and DP). We rationalize this by proposing a novel stochastic model based on a continuous time random walk (CTRW) approach that captures the observed intermittent behavior incorporating pore-scale structural and flow properties to bridge the pore-scale dynamics with macroscopic transport phenomena.

\section*{Results}

\noindent 
\textbf{Porous structure design and characterization}. 
We design a porous medium with a broad pore size distribution resembling the structure of natural soils or rocks. The medium structure (shown in Fig.~\ref{fig1}~$a$) consists of non-overlapping disks representing solid and impermeable grains (Fig.~\ref{fig1}~$b$) having random radius $r \in [0.06~\textrm{mm} < r < 0.12 \textrm{mm}]$ distributed as $P_r(r) \sim r^{-3}$, also shown in Fig.~\ref{fig1}~$c$. The distance between neighboring grains $\xi$ is power law distributed with an exponential cutoff,
\begin{equation}\label{xiPDF}
 	p_\xi(\xi) \sim \xi^{-\beta} e^{-\xi / \xi_0},
\end{equation}
where $\beta = 0.6$, $\xi_0 = 0.063$~mm, shown in Fig.~\ref{fig1}~$d$. The set of nearest neighbor grains is defined via a Delaunay triangulation~\cite{deAnnaPRF2017}. These grains of random size and location are placed within a folded strip of width $w = 3$~mm, equivalent to 48 times the characteristic pore size $\xi_0$ and length $L = 990$~mm, equivalent to $15,720 \, \xi_0$ (see Fig.~\ref{fig1}~$a$). We use soft lithography (see Methods) to print this geometry into a microfluidics chip (Fig.~\ref{fig1}~$a$).
\begin{figure*}[htb!]
	\centering
	\includegraphics[width=\textwidth]{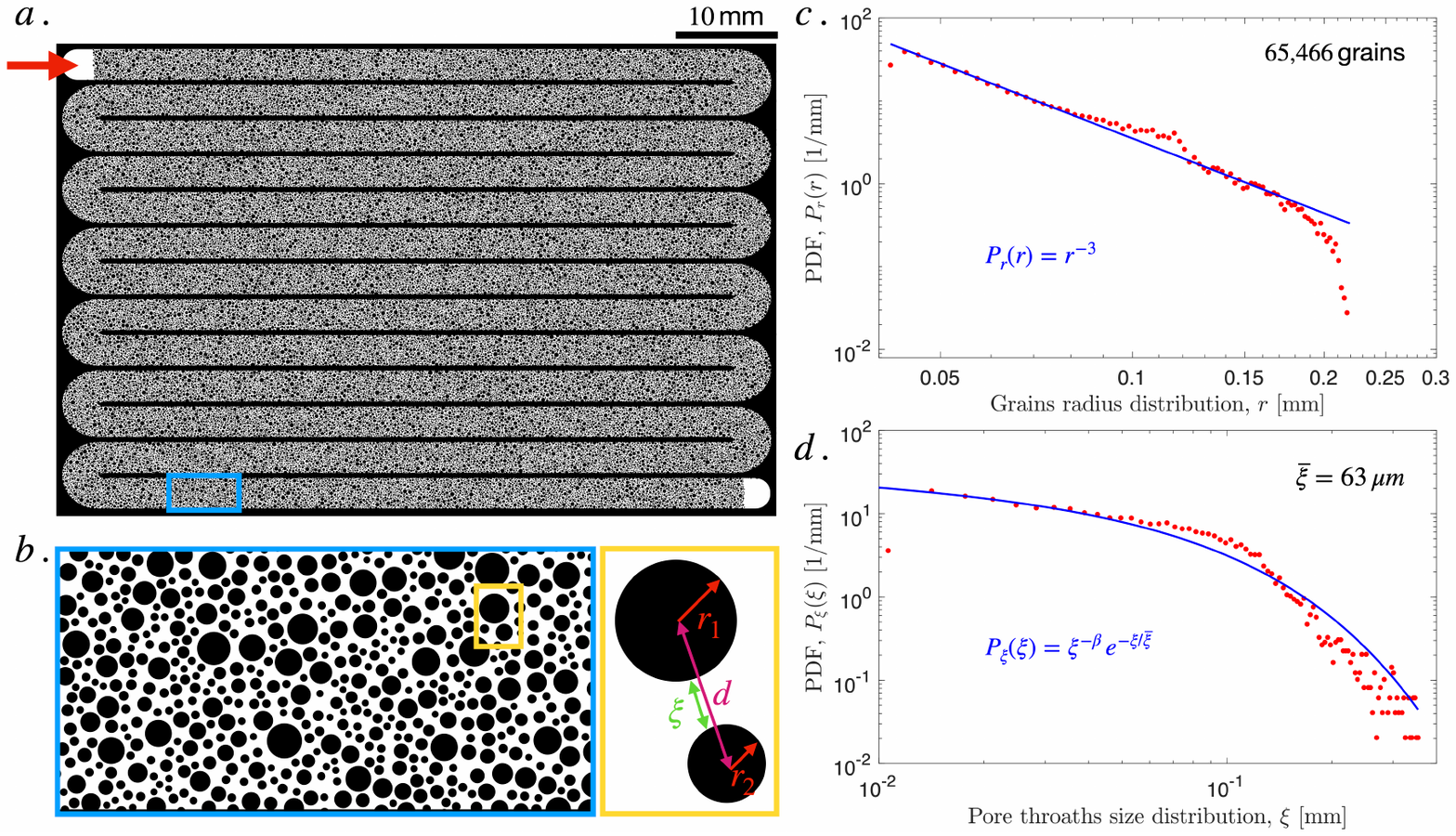}
	\caption{\textbf{Multi-scale microfluidic experimental set-up for filtration}. $a$) The porous structure is comprised of more than 65,000 grains: the longitudinal distance between inlet (top left) and outlet (bottom right) is $L = 970$~mm. $b$) A close-up view of the designed structure where solid grains are shown as black disks and pore throats are the space among closest neighbors. $c$) Statistical distribution (PDF) of grain radii $r$ and $d$) of pore throat size, $\xi$. }
	\label{fig1}
\end{figure*}

\noindent
To avoid the Hele-Shaw approximation for the flow within pores (that would result in a flat velocity profile among grains), we fabricate the described microfluidics devices with thickness $h = 0.05$~mm that is similar to the average pore throat $\xi_0$. The fluid moves from pore to pore passing through each individual throat of distributed size $\xi$: the local flow is well approximated by the Hagen-Poiseuille law, characterized by a parabolic profile. The designed structure is, thus, expected to control the broad power law distribution of low Eulerian velocities, $p_v(v) \sim v^{-\beta/2}$~\cite{deAnnaPRF2017}.\\

\begin{figure}[htb!]
	\centering
    \includegraphics[width=\textwidth]{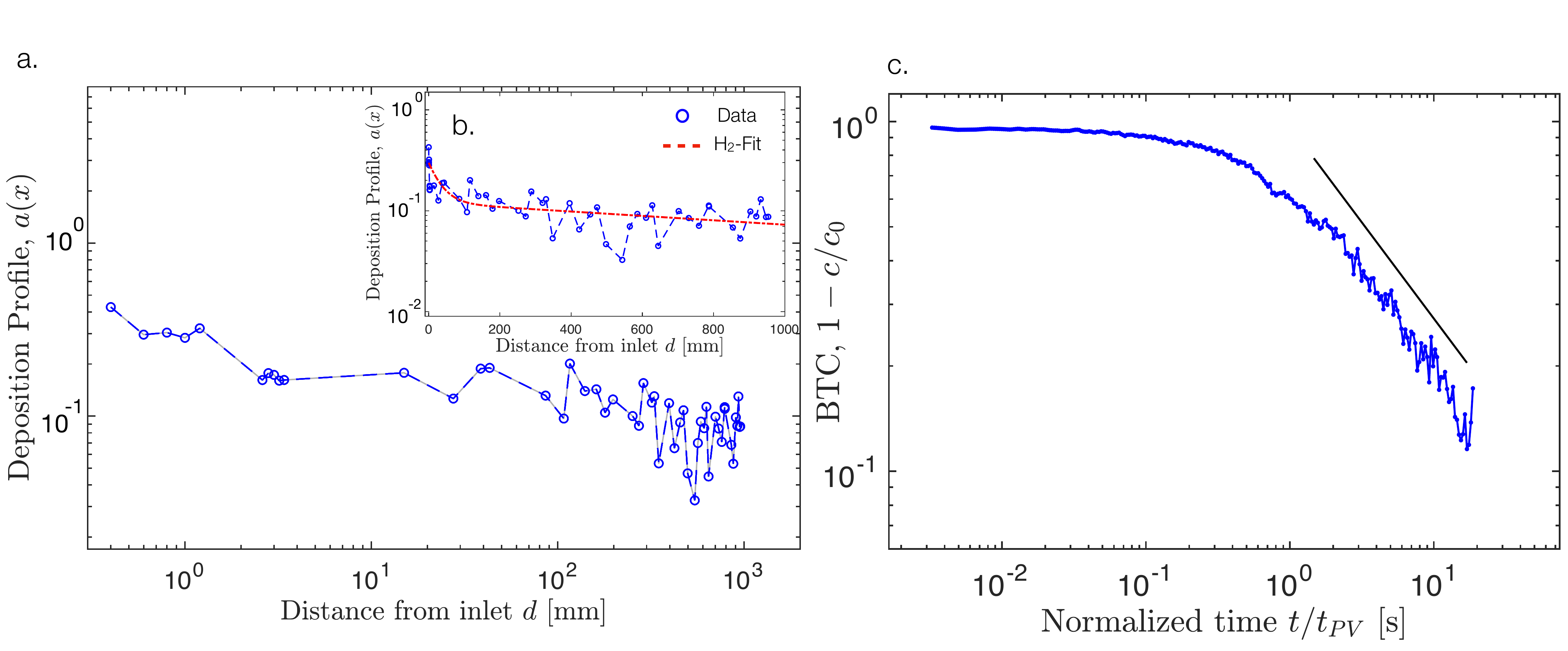}
	\caption{\textbf{Experimental colloidal transport and filtration by the porous structure.} Deposition Profile (DP) in panel $a.$) as the measured concentration of colloids attached to the medium solid surfaces along the distance from the inlet after $25h$ of continuous injection, $b$) Hyper-exponential fit $H_2(p,k_1,k_2)$ with $p=0.6$ and rates $k_1=0.03$ and $k_2=0.0005$.c.) Measured suspended colloids concentration at the medium outlet, Breakthrough Curve (BTC). Time is rescaled by the time needed to flush 1 pore volume $t_{PV}$. After $1t_{PV}$, a typical heavy tailing of the BTC signal with slope $t^{-0.67}$ is a signature of the anomalous transport, not captured by continuum approaches as prescribed by CFT theory.}
	\label{fig2}
\end{figure}

\noindent
\textbf{Macroscopic transport and filtration experiment}. 
We first saturate the microfluidic chip with a 1:1 milliQ water and D$_2$O mixture. The flow is imposed and controlled with a Harvard Apparatus PHD-ULTRA syringe pump, which is set at a constant flow rate $Q = 0.72 \, \mu$L/min. It results in an average (Darcy) velocity of $q^{\ast} = Q/A \phi = 0.16$~mm/s, where $A = h w = 0.15$~mm$^2$ is the chip cross-sectional surface and $\phi = 0.46$ is the medium porosity. Once the chip is saturated, we continuously inject a colloidal suspension for about 30 hours (corresponding to the elution of 20~pore volumes). The colloidal suspension is composed of polystyrene micro-spheres (diameter 1.1~$\mu$m and density $\rho = 1.05$~g/mL Thermofischer Fluoromax B100) dispersed in a density-matched 1:1 milliQ water and D$_2$O mixture, avoiding particle sedimentation. The relative importance of advection and diffusion processes is quantified by the corresponding Péclet number, $Pe = \overline{\xi} q^{\ast} / (\phi D) \sim 10^5$ based on the mean fluid velocity $q^{\ast}$ and the colloidal diffusion coefficient $D = 1.4 × 10^{-7}$~mm$^2$/s~\cite{bordoloi_structure_2022}. In such conditions, the flow is laminar since $Re = \xi q^{\ast} / \nu \sim 0.01$, where $\nu = 1$~mm$^2$/s is the water kinematic viscosity. In the following, time is normalized by the characteristic time necessary to elute a pore volume with the average Darcy velocity $q^{\ast}$, $t_{PV} = L/q^{\ast} = 103$~minutes.\\
\noindent
By comparing consecutive pictures collected with time-lapse video-microscopy of the fluorescent colloids within the pore space, we count the number of deposited and suspended colloids across the microfluidic reactor (as described in the Methods section). The deposition profile of colloids retained by the porous medium along the longitudinal position $x$, $a(x)$ after eluting 20 pore volumes is shown in Fig.~\ref{fig2}~$a$. It exhibits a non-exponential behavior, as shown in Fig.~\ref{fig2}~$b$ as blue dots. An hyper-exponential fit of the form of $H_2(x) = pe^{-k_1x} + (1-p)e^{-k_2x}/A$ (red solid line), where $A$ is a normalization factor, characterized by two characteristic scales fits relatively well the data that, thus, deviate from the CFT prediction. Fig.~\ref{fig2}~$c$ shows the complementary BTC, $1-c(t)/c_0$: as expected, after $t/t_{PV} = 1$ it quickly decays when colloids reach the outlet. However, it deviates from the exponential decay predicted by the classical models based on dispersion~\cite{Dentz2011} since at times larger than $t/t_{PV}= 1$, a power-law like heavy tail emerges, as characteristic of anomalous transport~\cite{Berkowitz2006}. \\

\noindent
Similar results are reported for column experiments~\cite{Tufenkjy_breacking_CFT, Sheibe} performed under unfavorable conditions, as we do here, that hinders colloids to surface retention (due to low ionic strength (IS) and high zeta potential). These previous observations have been interpreted invoking secondary energy minimum~\cite{Bradford2003} and multiple attachment rates due to colloidal suspension instability or grain roughness~\cite{Johnson2018}. It is important to note that all these aspects have no ground for the experimental conditions adopted here. The stability of the colloidal suspension is ensured by the very low IS value (0.5 mM): the suspension stays mono-dispersed during the entire experiment. Moreover, the grains are smooth cylindrical pillars with no contact points (no physical straining possible) and surface roughness much below the colloidal size~\cite{rusconi_microfluidics_2014}. \\

\noindent
Our observations suggest that new mechanisms, other than colloidal-surface interactions, have to be considered to understand transport in a heterogeneous medium. To move a step forward, we challenge the concept of a single collector \cite{Happel_1958} as the representative elementary unit for particles transport and filtration. In that context, the frequency of collisions is derived by considering the average flow velocity and the ratio between the radius of colloids and the radius of the average grain. \\ 

\begin{figure}[htb!]
	\centering
	\includegraphics[width=\textwidth]{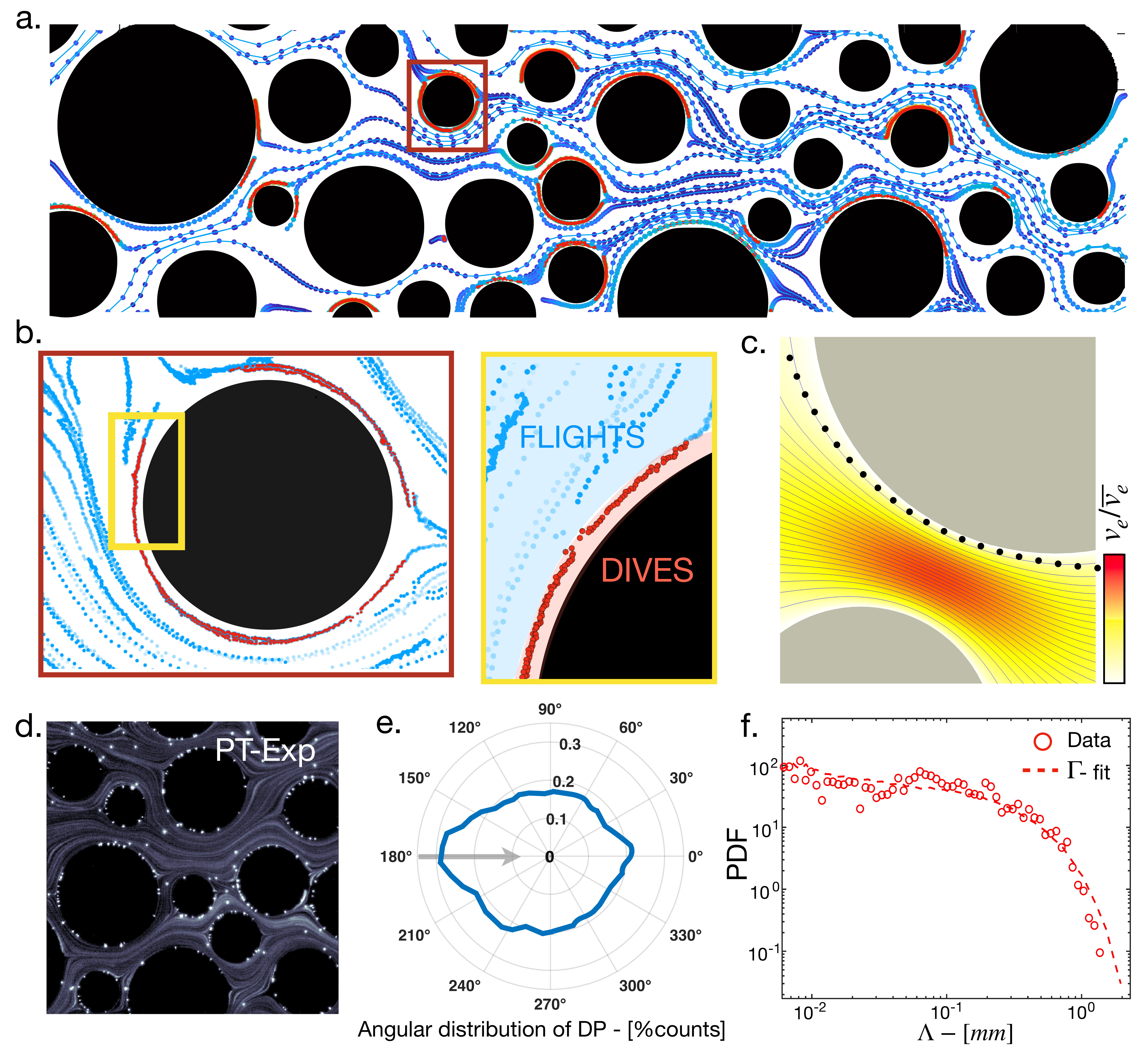}
	\caption{\textbf{Pore scale trajectories display flights-dives dynamics}. $a$) Reconstructed colloid trajectories among the grains through experimental particle tracking (PT). Each trajectory is decomposed into consecutive segments (highlighted in panel $b$) where particles are far (blue dots, flights) or close (red dots, dives) from a grain wall (zoomed view in yellow panel). $c$) Two-dimensional numerical simulation shows the hydrodynamic effect on purely advective streamline thickening: grain occlusion impacts the local velocity field ($v_e$) and leads to streamline compression with peaks in the pore throat, at the shortest grain-to-grain distance. The hydrodynamic pressure is responsible for deviations from pure parallel-to-grain surface direction, for a streamline falling into a dive (black dots). $d$) Time-lapse image stack of fluorescent colloids from PT experiment: blurry white stripes correspond to suspended particles while localized white spots are deposited particles that randomly surround each grain (black circles). $e$) Angular distribution of deposited particles computed around each grain for 10 different positions along the whole DP measurement: we observe a relatively even distribution, according to the local flow direction (in gray arrow), with slightly more attachment occurring upstream. $f$) Measured flights distribution, $p(\Lambda)$, from the PT experiments performed over 3~FoVs ($\approx 6$~mm). The dashed line shows the excellent fit of our model derived in SI~eq. (17): the cutoff is the FoV size.}
	\label{fig3}
\end{figure}

\noindent
\textbf{Hydrodynamic- and structure- induced filtration}.
To understand the drivers of macroscopic filtration by a heterogeneous porous system, we focus on the individual colloid trajectories reconstructed within 3 consecutive FoVs (about $6$~mm). Fig.~\ref{fig3}~$a$ shows a set of trajectories measured with Particle Tracking (PT) experiments in the same microfluidic device (see Methods). Remarkably, each trajectory is composed of a sequence of alternating segments along which the suspended colloids are near (red segments in Fig.~\ref{fig3}~$a,b$) or far (blue segments in Fig.~\ref{fig3}~$a,b$) from the solid grain walls, named dive and flight, respectively. Within a flight, trajectory segment farther than $d_c$, there are no physical conditions for the colloid to attach to a solid surface, as it travels too far from a grain to experience any surface interaction. It is, thus, simply transported by the flowing fluid. \\ 

\noindent
When a colloid falls into a dive-trajectory segment, closer than $d_c$ from a grain wall, it experiences the presence of a solid surface as a result of the balance between Van der Waals and electrostatic interactions. When colloids are suspended in pure water (i.e., with very low ionic strength) a strong repulsive electrostatic force emerges around each grain, raising the energy needed to approach the grain and experience attractive Van der Waals force. Under these unfavorable conditions (devoid of favorable heterodomains due to the smoothness of PDMS surface), pure diffusion cannot lead to attachment since the energy associated to Brownian motion is of the order of $k_BT$, while the repulsive energy barrier computed from the measured zeta potentials is of the order of several hundreds of $k_BT$. In fact, for our experimental setting, we measured the suspension zeta potential to be about $66$~mV, corresponding to an energy barrier of about $300 \, k_BT$ (see Methods and SI Fig.~6). Thus, another process must support the colloids in overcoming these barriers making attachment possible within a dive. Hydrodynamic forces play that role. \\

\noindent
Attachment of a particle can occur under unfavorable conditions when the net contribution of hydrodynamics, electrostatic repulsion and DLVO forces is positive towards a grain surface. This is possible only when trajectories are not parallel to solid surfaces. The curved shape of grains together with the proximity of other ones induces local deformation, i.e., compression and divergence, of streamlines leading to a hydrodynamic force pushing a particle toward a grain surface. This mechanism, also known as physical interception has been investigated previously~\cite{Herve2015,pradel_EST_2024}. To assess the streamlines of particles very close to a grain wall we exploited numerical simulations that overcome experimental limitations. Fig.~\ref{fig3}~$c$ shows trajectories from a high-resolution PT simulation performed in similar flow conditions (i.e. high \textit{Pe} with negligible diffusion), within a similar porous structure (PT simulation computed over velocity field from~\cite{deAnnaPRF2017}). Trajectories are never parallel to a grain surface even when very close (see Fig.~\ref{fig3}~$b,c$). Thus, hydrodynamic forces can push a colloid against the grain surface, making flow-induced physical interception possible. \\

\noindent
Previous works on physical interception around single collectors showed that colloid attachment happens at specific locations around each grain, where particle trajectories get compressed, like the closest point to the grain for the dotted trajectory in Fig.~\ref{fig3}~$c$. However, as shown in Fig.~\ref{fig3}~$d$, within a heterogeneous porous structure, physical interception takes place all around each grain. Single deposited fluorescent particles (white spots) surround grains without any preferential hot-spot for attachment. We extend this observation to the whole porous system by analyzing the statistical angular distribution of deposited particles around all grains at locations where the deposition profile was acquired (see Methods), as shown in Fig.~\ref{fig3}~$e$. We observe a relatively even distribution around grains (averaging across a few thousands), with slightly more attachment occurring upstream. Thus, we argue that it is reasonable to approximate the whole diving region (where attachment can take place) as the contouring area of each grain within a distance $d_c$ from its surface. \\

\noindent
Fig.~\ref{fig3}~$f$ shows the statistical distribution of the flight length $\lambda$ measured (red dots) from the PT experiment for an arbitrary critical distance of $d_c = 2 \, \mu$m. This distribution spans over more than 2 orders of magnitude. The analysis of Lagrangian flight length $\lambda$ shows that a significant number of particles cross the entire experimental field of view (FoV) without getting attached. Thus, the suspended and flowing colloids cross broadly distributed distances undisturbed (with no chance to attach for several pores), rather than surviving a multitude of collisions with each grain, typically seen as a single collector~\cite{YaoEST1971,Johnson2018}, until they enter a dive where the chance for removal from the suspension by interception occurs.\\   

\noindent
The three major results highlighted from the PT analysis are: i) each trajectory is composed of flights and dives with ii) a broad flight length distribution, and iii) a random attachment location around the grains. Based on these observations, we use a continuous time random walk (CTRW) framework to derive a stochastic model that reflects the intermittent nature of trajectories and predicts the observed macroscopic filtration. \\

\noindent
\textbf{Bridging pore-scale transport and macroscopic filtration}. Due to the stability and dilution of the colloidal suspension, we neglect interactions between the colloids. During the experiment duration, we injected a total of $2.8 \cdot 10^7$ colloidal particle (see Methods). Thus, we model the whole suspension with a similar number, $7 \cdot 10^6$, of particles that are simulated independently. Each trajectory is considered a 1-D sequence of segments where flights and dives occur alternately. Starting in a flight, during the $i-th$ step a colloid moves a random length $\lambda_i$ with random velocity $q_i$, which are independently distributed as $p_\lambda(\lambda)$ and $p_q(q)$ and are controlled by the medium structure. Each flight ends at location $x_i = x_{i-1} + \lambda_{i-1}$ and time $t_i = t_{i-1} + \frac{\lambda_{i-1}}{q_{i-1}}$, corresponding to the new dive starting position and time. Dives have been defined as the trajectory segments where a particle moves at a distance shorter than $d_c$ from the surface of a grain with random radius $r_i$. Diffusion and physical interception are modeled as stochastic processes. The particle crosses half the grain perimeter of random length $l_i = \pi r_i$ with a random velocity $u_i$, which are independently distributed as $p_l(l)$ and $p_u(u)$. These distributions are obtained from the distributions of grain size and throat length. The combined action of physical interception and diffusion is lumped into an effective attachment rate $k$, whose inverse defines the survival time $t_S = 1/k$. This rate reflects the macroscopic filtration efficiency that control the total amount of retained colloids (12\%, see Fig.~\ref{fig2}~$c$). \\

\noindent
If the transport time $l_i /u_i$ around a grain is shorter than the survival time $t_S$, the particle will escape the dive at location $x_i = x_{i} + l_{i}$ and time $t_i =t_{i} + \frac{l_{i}}{u_{i}}$, thereby transitioning towards the next step $i+1$ in a new flight. Otherwise, the particle attaches to the grain, its trajectory ends at location $x_i = x_i + u_{i} t_S$ at time $t_i = t_{i} + t_S$. This stochastic model constitutes a CTRW characterized by two alternating phases, termed alternated CTRW (A-CTRW) in the following. It is summarized as,
\begin{equation}
\begin{split}
 \left\{ \begin{array}{llll}
	x_i & = x_{i-1} + \lambda_{i-1} \\
	t_i & = t_{i-1} + \frac{\lambda_{i-1}}{q_{i-1}};\\
	t_S & = \infty \\
\end{array} \right.
\left\{ \begin{array}{ll}
	x_i & = x_{i} + l_{i} \\
	t_i & = t_{i} + \frac{l_{i}}{u_{i}};\\
	t_S & = 1/k_i
\end{array} \right.
\end{split}
\label{eq:CTRW}
\end{equation}
We compute the macroscopic DP and BTC from the distribution of the attachment locations and the distribution of the arrival time at $x = L$ of the particles surviving until the end of the domain, respectively. The segmentation of the whole filtration process into a sequence of piecewise steps provides the bridge between the microscopic structural- and flow- properties and macroscopic colloid transport. \\

\noindent
The BTC is measured as the distribution of colloid arrival times at the outlet given by the sum of all travel times of flights and dives crossed ($BTC(t)=\langle \sum_n t_{F_n} + t_{D_n}\rangle$, $n$ being the stochastic number of steps needed to reach the outlet for a trajectory). The tailing observed in Fig.~\ref{fig2} (from experiments) is a signature of the distribution of pore sizes combined with a Poiseuille flow profile within each pore. Thus, pore size variability controls the distribution of low velocities which enhances the dispersion of an injected plume of colloids, resulting in heavy tails in the arrival time distributions. The analytical description of the DP emerges from the interaction between transport and deposition~\cite{miele_stochastic_2019}. \\ 

\noindent
\textbf{Physical parameters}.
The proposed framework requires the definition of all physical quantities appearing in the A-CTRW model: the length and velocity distributions of flights ($\lambda,q$) and dives ($l,u$), reflecting the spatial variability of pore size and fluid velocity, and the constant attachment rate ($k$). \\

\noindent
\textbf{Distribution of dive jump sizes $l$}. As pointed out in the PT analysis, collision by physical interception is point-wise within each dive and occurs on specific streamlines. However, all particles that fall within a threshold distance $d_c$ from grain's surface can potentially diffuse and jump on such streamlines. Thus, it is natural to assume that the dive length is equal to the semi-perimeter of a grain, $l = \pi r$. This implies that $p_l(l) \propto p_r(l/\pi)$.  \\
\begin{figure}[htb!]
	\centering
	\includegraphics[width=\textwidth]{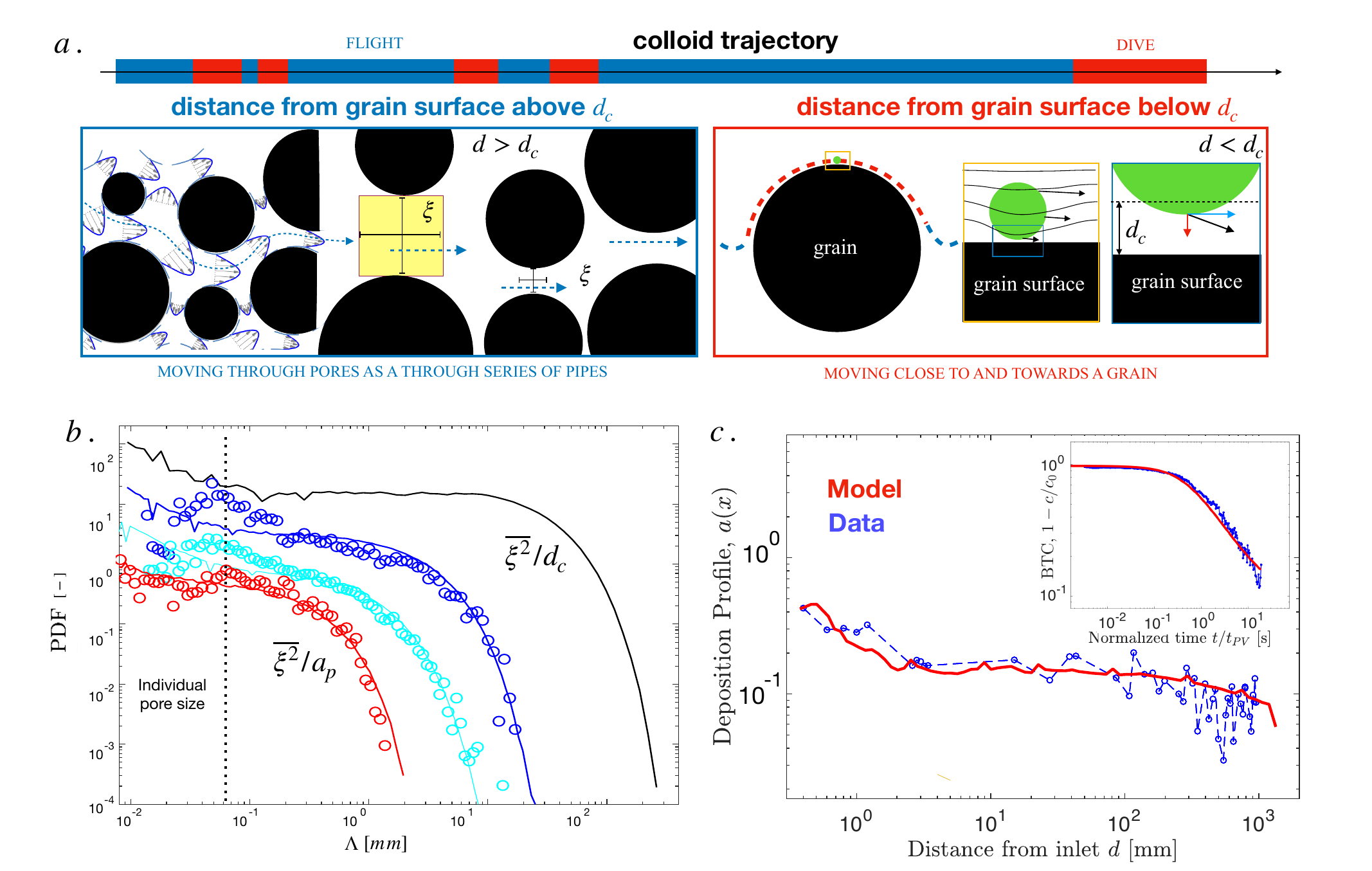}
	\caption{\textbf{Alternated-CTRW model for filtration}. a) Schematic representation of the one-dimensional Alternated-Continuous Time Random Walk model based on a sequence of flights (blue segments), where transport is dominated by random advection moving from pore to pore, and dives (red segments), where transport is dominated by motion along a grain wall surface where filtration can take place by physical interception. In a flight (blue inset), the motion within pores is represented as a sequence of squared pipe flow (yellow squares) where the parabolic profile is assumed to occur for the entire pore, neglecting thus edge's deformation due to grain's curvature~\cite{deAnnaPRF}. In a dive (red inset), the streamline gets deformed by local hydrodynamic pressure imposing a non zero component on the particle's velocity in the direction perpendicular to the grain wall. b) The Probability Density Function, PDF, of the flights' length, $\lambda$. Short flights, characterized by a single pore, have a power-law decay. Larger flight are dominated by flow channeling resulting in a flat flights distribution due to the equiprobable nature of longer flights. Larger flight jumps are characterized by an exponential truncation with cutoff of $\overline{\xi}^2 / d_c$ (see S.I. for full derivation, eq.~(17)). Experimental results are shown in red for the arbitrary $d_c = 1 \mu m$, direct numerical simulation for $d_c = 1 \mu m$ in cyan and for $d_c = 0.5 \mu m$ in blue, while prediction of flight distribution for $d_c = 0.01 \mu$m, peak of the energy profile shown in Fig.~SI~6 derived from DLVO theory, in black. c) Excellent fit of our A-CTRW (red solid line) to the experimental data (blue line) for DEP and BTC (inset).}
	\label{fig4}
\end{figure}

\noindent
\textbf{Distribution of dive velocity $u$}.
The fluid velocity within each pore throat is characterized by a parabolic profile \cite{deAnnaPRF2017,DentzJFM2018}, with the maximum velocity achieved at its center $u_M \sim \xi^2$. Thus, the velocity of a suspended particle moving at a fixed distance $d_c$ from the grain surface is given by the parabolic profile evaluated at $y = \xi/2 - d_c$, as $u = u_M[1 - (2 y/ \xi)^2] \sim \xi \, d_c$. In this way, the distribution of dive velocities $u$ is controlled by the pore size $\xi$ distribution and can be derived by variable transformation. It results in $p_u(u) = u^{-\beta/2}$, that is given by the porous structure consistently with the porelet model~\cite{deAnnaPRF} for low velocities.\\

\noindent
\textbf{Distribution of flight lengths $\lambda$}. Flights depend only on the statistics of transport. For a particle $p$, each flight length $\lambda$ is given by the sum of all pore sizes crossed during that phase, before entering a dive. Here, we assume that the length $\xi$ of each pore is statistically proportional to the throat size, as shown for similar structures~\cite{deAnnaPRF2017,jiao_PRF2024}: that is, long pores are also wide (see schematic in Fig.~\ref{fig4}~$a$). If within a flight a particle $p$ crosses $n$ pores and then approaches a grain at a distance $d < d_c$ from its surface, it transitions into a dive and the flight ends. Therefore, the flight length $\lambda$ is estimated as:
\begin{align}\label{eq:ell}
	\lambda = \sum\limits_{i = 1}^{n} \xi_i. 
\end{align}
We write the probability of a flight to have a jump length $\lambda$ as (see Supp. Info. for full derivation):
\begin{align}\label{lambdaeq}
	p_\lambda(\lambda) = \left\langle \sum\limits_{n = 0}^\infty \delta \left(\lambda - \sum\limits_{i = 1}^{n} \xi_i\right) \prod\limits_{i = 1}^n \alpha_i (1 - \alpha_{n+1}) \right\rangle
\end{align}
where $\alpha_i$ represents the probability that the particle does not dive within the $i$-th pore, and the average is intended over the pore size $\xi$, due to the random arrangement of grains. We assume that every time a colloid moves from one pore to another, it has a uniform probability of being at a distance $d$ from a grain wall. Thus, viewing a pore as a pipe, the probability of entering into a dive is given by the ratio between the area of a dive, an annular ring of thickness $d_c$ and diameter equal to $\xi / 2$, and the whole pore area, which is $\pi \xi^2 / 4$. This probability can be considered as an equivalent to the well known collector efficiency in CFT. While this parameter is classically assumed as a constant, it is here a stochastic variable that accounts for the randomness of each collector size. For the medium under consideration here, we obtain a closed expression for the distribution of flight length in the Laplace space, as discussed in the Supp. Info. \\

\noindent
Our analysis shows that the distribution of flight lengths is characterized by two main regimes, depending on whether the colloids experience $n = 1$ or $n \gg 1$ transitions. Indeed, for $n=1$, flights sample a single pore, so the  $p_\lambda(\lambda)$ follows the pore size distribution. For a larger number of transitions, the flight length distribution collapses to a flat profile with a characteristic exponential decay of characteristic scale $\xi_0^2 / 2d_c$ (see Supp. Info.). \\

\noindent
Directly measuring the distribution of the flight lengths is particularly challenging for two reasons. First, flights span across several length scales. As pointed out previously, a significant number of colloids crossed undisturbed the entire distance domain of the FoVs ($6mm$ in total, corresponding to about 100 average pore sizes) in the PT experiment. This means that to properly track and sample longer flights, a larger domain for PT is needed. Although we achieved this by performing multiple PT experiments on 3 consecutive FoVs, we experimentally noted that the number of reconstructed Lagrangian trajectories gets sensitively reduced proportionally to their total length. Second, performing particle tracking within a heterogeneous porous system requires a high acquisition rate (we used $60$ Hz), which limits the time duration of the observation collected. Moreover, to measure $p_\lambda(\lambda)$, we impose the arbitrary $d_c = 2\,\mu$m, which is larger than the actual critical distance below which attachment is possible ($d_c = 0.01\, \mu$m, see Supp. Info.) but allows us to collect enough statistics. The measured distribution is reported in Fig.~\ref{fig4}~$b$ as red circles together with the Monte-Carlo simulation of eq.~\eqref{lambdaeq} (red line in Fig.~\ref{fig4}~$b$), fed only with the designed pore size distribution $p_\xi(\xi)$.  \\

\noindent
To further validate this result, we performed high-resolution 2D PT simulations, as in~\cite{deAnnaPRF2017}, on a similar porous geometry (see Methods) but covering a much larger amount of space, about $10^3$ average pore sizes ($10^2$ for the experiment). Then, we analyze these simulated trajectories with two values of the critical distance: $d_{c1} = 1 \, \mu$m and $d_{c2} = 0.5 \, \mu$m. The measured flights length distribution for the simulated trajectory are shown in Fig.~\ref{fig4}~$b$ (cyan symbols for $d_{c1}$ ad blue symbols for $d_{c2}$). The model of eq.~\eqref{lambdaeq} fits very well the simulated data, just considering the pore size distribution and the critical distances $d_{c1}$ ad $d_{c2}$. \\

\noindent
To predict the macroscopic deposition profile, we use our model prediction with the designed pore size distribution and $d_c = 0.01 \, \mu$m, which corresponds to the location of the energy barrier peak predicted by DLVO for the measured $\zeta$-potential ($V_\zeta = - 66$~mV)~\cite{kirby_zeta_2004,yu_evaporative_2017}.\\

\noindent
\textbf{Distribution of flight velocities $q$}. The velocity $q$ of a flight is given by $q = \lambda / \tau_F$, where $\tau_F$ is the flight duration. Thus, to estimate the $q$ distribution, we, first, estimate the number of pores composing the average flight as $n_p = \lambda / \xi$ which is controlled by the distributions of $\lambda$ (that we measured) and $\xi$ (imposed by the porous structure). Then, the flight velocity is $q = \lambda / \tau_F = \lambda / (\sum_{i=1}^{n_p} \xi_i / v_i)$, where the pore size $\xi$ distribution is given by the medium structure and the individual pore velocity $v$ distribution can be estimated by employing the porelet model~\cite{deAnnaPRF2017}.\\

\noindent
\textbf{Attachment rate $k$}.
The value of the attachment rate $k$ is estimated by matching the ratio of recovered mass and injected mass from the BTC signal, estimated here by $k = 4.3 \cdot 10^{-4}$~s$^{-1}$ corresponding to a survival time $t_S = 1 / k = 2325$~s or about 39 minutes.\\

\noindent
\textbf{Model prediction}. Fig.~\ref{fig4}~$c$ shows the excellent agreement between the proposed model (red line) for deposition profile $a(x)$ and the complementary BTC $1-c(t)/c_0$ (inset) measured with our microfluidic setup (blue dotted lines). The model we developed has two major outcomes. On the one hand, it properly predicts the dynamics of the transported colloids that turns out to be controlled by the pore throat size distribution, $p_\xi(\xi)$, that shapes the late arrival times of the BTC. On the other hand, the spatial organization of the retained colloids has a two regimes: power-law near the inlet and exponential decay at larger distances. This emphasizes the role of flights length statistical distribution which is  controlled by the porous structure and its physico-chemical properties in terms of the pore size PDF $p_\xi(\xi)$ and the critical distance $d_c$.

\section*{Discussion}
\begin{figure}[htb!]
	\centering
	\includegraphics[width=17cm]{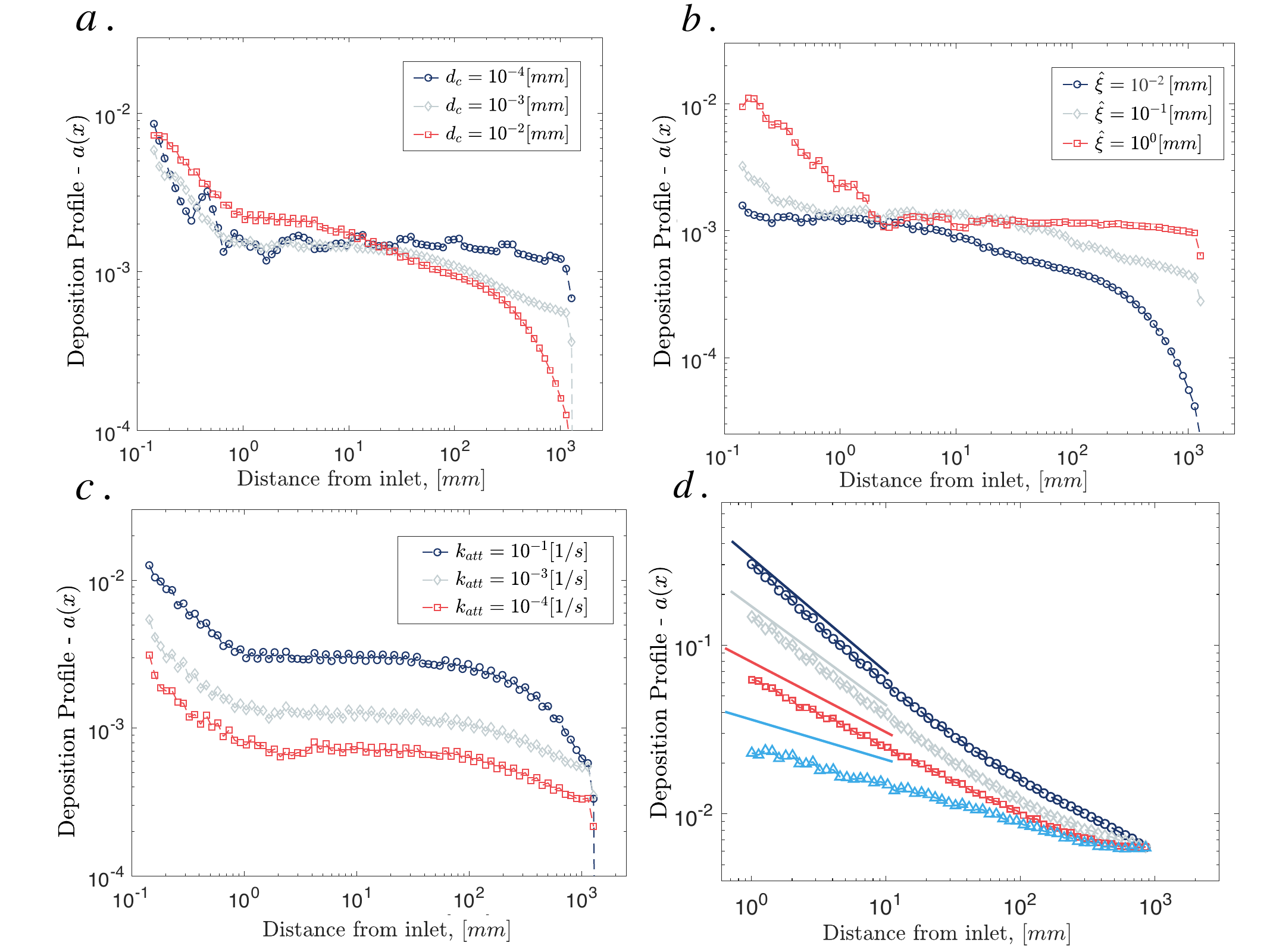}
	\caption{\textbf{Parameters controlling the deposition profile}. The dependence of DP by the microscale parameters as predicted by the derivation of flights distributions and eq.~\eqref{eq:cutoff} is exploited in the following scenarios. Starting from the same parameters of the experimental set-up, we systematically varied $a$) the critical distance $d_c$ that directly affects the exponential cut-off of the flights jump distribution $\lambda$; $b$) the pore size cut-off $\xi_0$; $c$) the average attachment rate; and $d$) the characteristic pore size distribution exponent $\beta$: the solid lines represent $x^{-\beta}$.}
	\label{fig5}
\end{figure}

\noindent
The proposed A-CTRW shows that the dynamics (BTC) and the spatial organization (DP) of macroscopic filtration are shaped by the microscopic pore structure through three key quantities: i) the critical distance $d_c$, that depends on the material properties of the colloidal suspension and porous system, ii) the medium structure, in terms of pore size distribution characterized by its cut-off scale $\xi_0$ and scaling exponent $\beta$, and iii) the attachment rate $k$. \\

\noindent
The designed experiment mimics the heterogeneity of a natural soil  structure, controlling the grain and pore size distribution across almost 2 orders of magnitude, making this setup suitable for microfluidics replication at the expense of limited reproduction of even more heterogeneous structures. Finer particulates, aggregates, heterogeneity in soil texture, and the presence of macro-fractures should be reflected in broader distributions. Dissecting the dynamics of filtration through the mechanism of flight and dives we observe that the DP is primarily controlled by the flight distribution and, thus, by $\xi_0$ and $\xi_0^2/(2d_c)$ for short and long distance, respectively. However, we note that the exponential cut-off for the deposited particles profile, $L_c$, is farther away compared to the cutoff of flight PDF, $\xi_0^2 / (2d_c)$. This is not unexpected, as the stochastic nature of adhesion implies that falling into a dive does not necessarily correspond to being attached. In the following, we discuss how the DP is controlled by the medium structure and its physico-chemical properties, specifically with respect of the parameters $\xi_0$, $\beta$, $d_c$ and $k$. \\

\noindent
\textbf{Pore size cut-off $\xi_0$ and critical distance $d_c$}.
Fig.~\ref{fig5}~$a$ shows the prediction of the A-CTRW model for the deposition profile varying the critical distance $d_c$. This parameter controls the location of the DP large distance cut-off $L_c$, which reasonably depends by two processes: the average flight size $\overline{\lambda}$ and the average number $n$ of flight-dive phases experienced by colloids before attaching, $L_c = n \overline{\lambda}$. Conceptually, this cut-off arises from the combination of advective transport and attachment. Flights are responsible for the advective component, and they are characterized by their own cutoff $\xi_0^2/(2d_c)$ which encodes the fact that eventually a particle will fall into a dive. However, falling into a dive does not necessarily mean that a particle gets attached. The average number $n$ of flight-dive phases experienced before attachment can be derived from the probability to survive $n$ phases, which is controlled by the ratio of the survival time near a grain, $t_S = 1/k$, and the transport time, with average velocity $v$ along the average grain size $\overline{r}$, $t_D = \overline{r} / \overline{u}$. Thus, $n = t_S / t_D = \overline{v}/\overline{r} \, k$. Finally, the cut-off for the deposition profile is given by
\begin{align}\label{eq:cutoff}
	L_c = n \overline{\lambda} = \frac{\overline{v}}{\overline{r}\, k} \, \frac{\xi_0^2}{2d_c}.
\end{align}
If $t_S / t_D = \frac{\overline{u}}{\overline{r}\, k} > 1$ the deposition profile cut-off would exceed the flight size cut-off, otherwise, the attachment rate would be so strong to significantly shorten the colloids ability to travel through the medium: only a few colloids would be able to remain in the suspended phase. Fig.~\ref{fig5}~$a$ shows that increasing the critical distance $d_c$ does not change the deposition profile for short distances, but since the overall retention of colloids is more frequent, the cut-off distance $L_c$ decreases as predicted by eq.~\eqref{eq:cutoff}. By increasing the cut-off pore size $\xi_0$, which corresponds to considering pores varying across a wider range of values, it extends the first regime of the DP, the one controlled by the pore size distribution. This is supported by the model predictions shown in Fig.~\ref{fig5}~$b$, for which the short distances deposition scales as the pore size distribution until $\xi_0$.\\

\noindent
We remark that eq.~\eqref{eq:cutoff} provides a further analogy with the classical definition of attachment rate, $k_{CFT} \propto \eta \alpha$, where $\eta$ describes the probability of collision, while $\alpha$ defines the probability of attachment, only if a collision occurred. Despite their empirical estimation in continuum theory, these quantities are here derived, in terms of process-based considerations.\\  

\noindent
\textbf{Attachment rate $k$}.
Chemical and surface properties for both, colloidal suspension and porous material, are lumped into the attachment rate value $k$: higher values will represent more favorable conditions. As discussed, the deposition profile is primarily controlled by the flight length distribution that depends on transport and structural properties of the porous system. We expect that changing the value of the attachment rate $k$ will not affect the shape of the deposition profile, but only the overall overall fraction of the retained particles. Thus, increasing $k$ will result in a upwards shift of the DP and a decrease of its long-distance cut-off. Our model prediction, shown in Fig.~\ref{fig5}~$c$, confirm this hypothesis. The scaling of the deposition profile does not change, it is shifted vertically and its large distance cut-off varies as predicted by eq.~\eqref{eq:cutoff}.\\

\noindent
\textbf{Scaling exponent $\beta$ of the pore size distribution}.
The scaling exponent $\beta$ describes the pore-size distribution for value smaller than the cut-off, $\xi < \xi_0$. In order to probe the impact of $\beta$ on DP, we fix the value for $\xi_0$ to be much larger than the spatial domain size, and we vary $\beta$. Fig.~\ref{fig5}~$d$ shows, as expected, how changing $\beta$ changes the deposition profile as $\sim x^{-\beta}$. \\

\noindent
\textbf{Outlook}. Due to the multi-scale structure of complex and heterogeneous porous systems, how colloids are transported through pores and deposited over grains has remained poorly understood. This lack of knowledge is due also to the lack of a theoretical frameworks able to bridge microscopic transport/retention and macroscopic filtration. Classical and current models rely on the assumption that particle removal is assumed to happen at a constant rate over defined spatial scales. However, such frameworks are constantly defied by experimental evidence, revealing the limits of such continuum approaches. The multiscale microfluidics model system we developed here allowed us to observe across spatial scales (from tens of microns to a meter) the transport and deposition of a mono-dispersed colloidal suspension constantly injected through a porous system of controlled heterogeneity. Colloids intermittently move along trajectories composed of alternating flights and dives, during which they cannot or can get attached, respectively. We rationalize this remarkable observation by proposing an alternated continuous time random walk (A-CTRW) that honors the observed intermittent behavior and bridges the microscopic medium structure and macroscopic spatial organization and dynamics of filtration. Due to the ubiquity of complex geometry and velocity disorder in natural and engineered porous media (such as soils, rocks and membranes) and the general nature of our model system, we anticipate that the fundamental mechanisms and their large scale manifestations uncovered here will apply to a broad range of transport problems within natural and engineered porous environments.

\noindent
\section{Methods}

\subsection*{Microfluidics fabrication}
\noindent
The designed grain structure was printed onto glass at super high resolution (JD Photodata) in chrome. Micro-channels were fabricated by prototyping against a silicon master (\emph{Electro-Technic} products) with positive relief features using standard soft lithography techniques. Liquid Polydimethylsiloxane (PDMS, Sylgard 184 Silicone Elastomer kit, Dow Corning, Midland, MI) was mixed with the curing agent provided by the Sylgard 184 kit of ratio 1:10 and placed at 60 \textdegree C for 5 hours to harden. \\

\noindent
The PDMS layer was 50~mm wide and 75~mm wide in order to fit onto a glass slide of same size. The channel (engraved into the PDMS layer) has thickness $h = 0.05$~mm (as the relief of the silicon master). We plasma-bond it to a soda-lime glass slide (75~mm long, 50~mm wide, and 1~mm thick): this plasma treatment yields the PDMS surface temporarily hydrophilic facilitating their subsequent saturation with the saturation medium. Prior to plasma bonding, the glass slide was coated with a thin layer (about 50~$\mu$m) of PDMS by spin-coating: thus, the total surface of the microfluidics channel to which colloids are exposed is composed of PDMS only (the suspension injected into the channel is never in contact with the glass). Thus the colloids-surfaces interaction is homogeneous, avoiding any bias in the attachment process between the PDMS walls and the glass. \\

\subsection*{Colloidal suspensions}
\noindent
We prepared a colloidal suspension composed by a 4~ml solution of $56 \%$~UV treated MilliQ water and $46\%$~D2O (Sigma-Aldrich) to which we add $6~\mu$l of the mono-disperse fluorescent polystyrene particles FLUOROMAX B0100 at concentration 1\% solid (Thermo Fisher Scientific) of $1.1 \, \mu$m diameter, $R = 0.55 \, \mu$m of radius. Heavy water is added to the suspension (1:1 mixture of MilliQ and heavy waters), so that we match the colloids density, which is 1.05 g/ml, to avoid their sedimentation. Before starting each flow experiment, the used suspension has been sonicated for 20 minutes in a bath with ultrasound, in order to breakdown eventual clusters into individual colloids. We measured the injected concentration $c_0$ analyzing pictures taken in the microfluidics region between the inlet hole and the first line of the grains. Finally, the BTC are measured as the number of colloids detected after a time $t$ of injection per unit volume $c(t)$, divided by the injected number of colloids per unit volume $c/c_0$. During the overall duration of the experiment ($T = 30$~hours), we injected at constant flow rate $Q = 0.72 \, \mu$l/min a volume $QT = 1.3$~ml of suspension. Since the volume of a single colloid is $4/3 \pi R^3$ and the FLUOROMAX suspension has concentration 1\% solid, this injected volume corresponds to a total number of about $2.8 \cdot 10^7$ particles. \\

\noindent
The ionic strength of the suspension was fixed to 0.5~mM by adding NaCl to the colloidal suspension. The Zeta Potential of the colloidal suspension ($\zeta = -65.6 \pm 0.6$ mV) has been measured with the Zetasizer Nano ZS (MALVERN). The interaction energy profile (see S.I., Fig.~S6) for the system PDMS-water-colloids is computed following the DLVO theory (\cite{ruckenstein_adsorption_1976, Hogg1966,Elimbook}, with $H_p$ ,$H_w$ and $H_{PDMS}$ as Hamaker constant for latex particles, water and PDMS, respectively (\cite{ruckenstein_adsorption_1976, Hogg1966,yu_evaporative_2017}and S.I. for details).

\subsection*{Time-lapse video-microscopy} 
\noindent
Images were acquired with an automated inverted microscope (Eclipse Ti2, Nikon) equipped with a sCMOS camera (Hamamatsu ORCA flash 4.0). Each picture, of 2048 by 2048 pixels, was recorded at $10X$~magnification (corresponding to 0.65~$\mu m$/pixel) was focused on the middle height of the microfluidic channel. For the BTC, one image was sufficient to cover enough outlet area, while for DP measurement, a grid of $4 \times 4$ images was acquired and stitched together by the software NIS Elements. For the BTC measurement, colloids were imaged with a fluorescence optical configuration, using a Nikon DAPI filter cube, with an exposure time of 50~ms. For the DP, the exposure time was three times longer to make deposited particles brighter than the suspended and help the subsequent detection process.\\

\subsection*{Particle Tracking experiments} 
\noindent
Our Particle Tracking (PT) is composed of two steps. First, microscopy time-lapse image acquisition at 100~Hz is performed at a given location, capturing a given FoV, with the same microscopy setting used for BTC measurements. Due to the high frequency in acquisition rate, the time window for continuous acquisition was limited at $40 sec$ (thus, 4,000 pictures). After that, a new acquisition started. This process is repeated 6 times in each single FoV, before moving to the next FoV, an area of the microfluidics down-stream and adjacent to it. We collected data for 3 adjacent FoV in total. Trajectories are, then, reconstructed by pairing assignment using a combination of PT algorithms that has been developed in~\cite{morales_stochastic_2017} and further tested in~\cite{berghouse_PT_SciRep2024}. The first round of ID assignment between detected particles has been done using the Kalman algorithm of the Trackmate plugin available in FIJI~\cite{schneider_nih_2012}. This resulted in truncated trajectories: we have glued those trajectories that match final and initial location and velocity. This is done using V-track code to obtain longer trajectories within the same FoV. The same algorithm is further used to glue trajectories between adjacent FoVs, also matching location and velocity. We refer to~\cite{berghouse_PT_SciRep2024} for all the details. To segment flights and dives, the exact location of the grains edge relative to the centroids of suspended particles, was assessed. This was achieved by reconstructing the grains mask where grains outlines were defined by imaging the microfluidics saturated with a fluorescent tracer.\\

\subsection*{Measuring Deposition Profiles} 
\noindent
Every hour, we acquired a series of pictures sampled in pairs, separated by a time lag $\Delta t_1 = 5$~minutes, along the longitudinal direction, at 40 locations (covering about 10~\% of the whole system). By comparing two pictures sampled at the same location with the time delay $\Delta t$, we assess which particles did not move and, thus, are attached to a solid surface. This results in the profile $a(x) = V_c(x) / V_p(x)$ defined as the ratio between the total volume of attached particles at a distance $x$ from the inlet ($V_c(x) = N_c(x)*4/3\pi R^3$) and the pore space volume detected in the images collected at that same location ($V_p(x)$), where $N_c$ is the total number of detected colloids, computed by exploiting the \emph{pkfnd} routine of \emph{Matlab} in each image $I_c$. The images location is not equally spaced: we sample more frequently the zone close to the inlet and gradually decrease the number of sampling points towards the outlet: in this way, we can plot our spatial measure with a logarithmic scale honoring every decade of the axis. \\ 

\noindent
The used polystyrene colloids of diameter $1.1 \, \mu$m that have a diffusion coefficient $D \sim 4 \cdot 10^{-7}$mm$^2$/s~\cite{bordoloi_structure_2022,hamada_method_2023}, thus over a time lag $t_a = 300$~s they will move randomly and explore an area of size $\sqrt{2Dt_a} = 15 \, \mu$m. We evaluate as very unlikely the event of a diffusing particle wondering about to come back at the exact same location after a time $t_a = 300$~s. Thus, we consider as not moving the particles that are found at the same location after 300 seconds, the time separating two acquired images. Attached particles correspond, thus, to the non zero elements of the matrix defined by the product $I_c = \sqrt{I_t \cdot I_{t+\Delta t}}$, where $I_t$ is the intensity captured by the picture at time $t$.\\

\noindent
In the resulting matrix $I_c$ are present several clusters of bright pixels over a dark background. These clusters have individual area that has a size given by the number of colloids that compose them. In our case, most clusters have an area equivalent to that one of a single colloid, but some have an area equivalent to few colloids (2 to 4). Only in a few cases these clusters are large and composed by tens of colloids. These cases happen because the attachment of a single colloid represent an hot-spot for attachment of new coming colloids. This has been reported~\cite{Herve2015} in microfluidics channels designed to have a single, or few, obstacle (single collector). We verified that we are, here, in the so-called \emph{clean bed assumption} corresponding to the case in which a flowing-suspended colloid will experience a solid grain wall which is mostly free of other attached colloids (multiple attachment events are very unlikely). We remark that removing clusters much larger than a few colloids do not impact the DP. Moreover, we confirm that considering only clusters attached to the surrounding area of each grains, instead of the whole chip, has no effect on the DP shape, but just induces a slight vertical shift. \\

\subsection*{Measuring BTC}
\noindent
We periodically acquire pictures in pair separated by a short time lag $\Delta t_2 = 2$~s every 5 minutes at the outlet $x = L$. The comparison of 2 consecutive pictures separated by $\Delta t_2$ allows us to detect particles that are moving and, thus, to measure the breakthrough curve (BTC) which is the concentration of the suspended colloid, normalized by the injected one, $c(t)/c_0$. Contrary to the image process for DP, suspended particles at time $t$ are here obtained by the non-zero elements of the matrix given by $I_t - \left( \sqrt{I_t \cdot I_{t+\Delta t}}\right)$. Also for the BTC the particle count has been done exploiting the \emph{pkfnd} routine of \emph{Matlab}: we took care of looking for clusters of pixels slightly larger than the one chosen for the deposited particles since moving particles leave a larger image, as previously discussed. Here, the flowing particles are detected as slightly elongated bright ellipsoids rather than circular objects due to the local velocity they are experiencing when the picture is collected. \\

\subsection*{Particle Tracking simulation}
\noindent
We integrate $50,000$ particles' trajectories by interpolating, with a higher order numerical Methods (see also~\cite{deAnnaPRF2017}), the velocity field $\boldsymbol{v}(\boldsymbol{x})$ at each particle location $\boldsymbol{x}_p$ to get the particle velocity $\boldsymbol{v}_p$ and integrate $d \boldsymbol{x}_p /dt = \boldsymbol{v}_p(\boldsymbol{x}_p)$. The simulated porous structure (different form the experimental one) is characterized by the pore throats size distribution $p_\xi(\xi) \sim \xi^{-\beta} \, e^{-\xi/\xi_0}$ with $\beta = 0.16$.\\

\noindent
\textbf{Additional information}\\
A detailed description of the mathematical derivation for the characteristic scaling laws can be found in the Supplementary Information. \\

\noindent
\textbf{Acknowledgments}\\
F.M. acknowledges the Swiss National Science Foundation - Postdoc Mobility Fellowship (SNF, grant Nr. P2LAP2$\_$199473). V.L.M. acknowledges the support of the U.S. National Science Foundation (EAR-1847689, EAR-2345366). P.d.A. acknowledges the support of FET-Open project NARCISO (ID: 828890) and the Swiss National Science Foundation (grants ID~200021$\_$172587 and 200021$\_$219863).

\noindent
\textbf{F.M. and P.d.A. designed the research, F.M., H.T. and P.d.A. F.M. designed the experimental set-up and procedure, F.M., A.D.B. P.d.A and V.L.M. designed and did Particle Tracking experiment, and F.M., A.D.B. and P.d.A. analyzed the data, F.M., V.L.M., M.D. and P.d.A. derived the theoretical model, and all authors wrote the manuscript.} \\

\noindent
\textbf{Competing financial interests}\\
The authors declare no competing financial interests.\\

\clearpage
\section*{Supplementary Information}
\addcontentsline{toc}{section}{Supplementary Information}
\setcounter{figure}{0}
\renewcommand{\thefigure}{S\arabic{figure}}

\setcounter{section}{0}
\renewcommand{\thesection}{S\arabic{section}}
\section{Derivation of Flight size PDF solution}

We consider a colloid that is moving between pores of different length $\lambda$. We assume that the length and the diameter are the same. That is, a long pore has also a large diameter. If the colloid enters the pore at a distance $R_c$ from the wall it dives. The probability that the colloid does not dive, that is, that is flies upon entering a pore is denoted by $\alpha_i$. The latter can depend on the properties of the pore, like its length and diameter. The probability $p_n$ that a particle flies exactly along $n$ pores before it dives is given by 
\begin{align}
\label{eq:pn}
p_n = \prod\limits_{i = 1}^n \alpha_i (1 - \alpha_{n+1}), 
\end{align}
that is, the probability to fly along $n$ pores times the probability to dive at the $n+1$th pore. Thus, the distribution of flight lengths is given by
\begin{align}
\label{eq:p}
p(x) = \left\langle \sum\limits_{n = 0}^\infty \rho_n(x) p_n \right\rangle, 
\end{align}
where $\rho_n(x)$ is the distribution of flight distances after $n$ steps,
\begin{align}
\rho_n (x) = \delta(x - \ell_n).
\end{align}
The flight distance $\ell_n$ after $n$ pores is
\begin{align}
\label{eq:ell}
\ell_n = \sum\limits_{i = 1}^n \lambda_i. 
\end{align}
The angular brackets denote the average over all colloids. We can write $p(x)$ by using~\eqref{eq:pn} and~\eqref{eq:ell} as
\begin{align}
p(x) = \left\langle \sum\limits_{n = 0}^\infty \delta\left(x - \sum\limits_{i = 1}^n \lambda_i\right) \prod\limits_{i = 1}^n \alpha_i
(1 - \alpha_{n+1}) \right\rangle
\end{align}
%

\subsection{Constant diving rate and constant pore-length}

We consider the first the case of constant diving rate $\alpha_0$ and pore length $\lambda_0$. Thus, we obtain for $p(x)$
\begin{align}
p(x) = \sum\limits_{n = 0}^\infty \delta\left(x  - n \lambda_0\right) \alpha_0^n
(1 - \alpha_0). 
\end{align}
That is, the probability $p_n$ of $x = n \lambda_0$ is
\begin{align}
p_n = \alpha_0^n (1 - \alpha_0) = \exp(n \ln \alpha_0) - \exp[(n + 1) \ln
\alpha_0] = \exp(- x_n/\lambda_c) - \exp(-x_{n+1}/\lambda_c) = H(x_n > \lambda_0),
\end{align}
where we defined $\lambda_c = - \lambda_0/\ln \alpha_0$ For $x \gg \lambda_c$, we can write
\begin{align}
p_n = \frac{\Delta x \exp(-x/\lambda_c)}{\lambda_c},
\end{align}
that is
\begin{align}
p(x) = \frac{p_n}{\Delta x} = \frac{\exp(-x/\lambda_c)}{\lambda_c}. 
\end{align}
Thus, it decays exponentially fast on the length scale $\lambda_c = - \lambda_0/\ln \alpha_0$.

\begin{figure}
\includegraphics[width=0.8\textwidth]{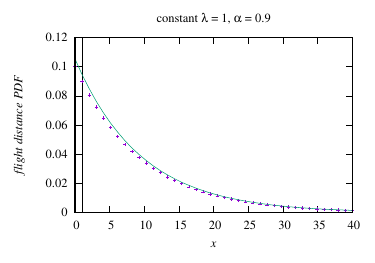}
\caption{Distribution of flight distances for constant pore length and diving rate.\label{fig1}}
\end{figure}

\subsection{Constant diving rate and exponential pore-length distribution}

We consider the case the $\alpha_i = \alpha_0 = $ constant and $\lambda_i$ exponentially distributed with characteristic scale $\ell_0$. In this case,
\begin{align}
\rho_n(x) = \left(\frac{x}{\ell_0}\right)^{n-1}
\frac{\exp(-x/\ell_0)}{\Gamma(n) \ell_0}. 
\end{align}
for $n > 0$. Thus, we obtain for the flight distance distribution
\begin{align}
p(x) = \frac{\exp(-x/\ell_0)}{\ell_0}(1 - \alpha_{0})  \sum\limits_{n =
  1}^\infty \left(\frac{x}{\ell_0}\right)^{n-1}
\frac{\alpha_0^{n}}{\Gamma(n)} + (1 - \alpha_0) \delta(x)
\end{align}
The sum can be written as 
\begin{align}
& \sum\limits_{n = 1}^\infty \left(\frac{x}{\ell_0}\right)^{n-1}
\frac{\alpha_0^{n}}{\Gamma(n)} = \left(\frac{x}{\ell_0}\right)^{-1} \sum\limits_{n =
  1}^\infty \left(\frac{x}{\ell_0}\right)^{n}
\frac{\alpha_0^{n}}{\Gamma(n)}
\\
&= \left(\frac{x}{\ell_0}\right)^{-1} \left[\sum\limits_{n =
  0}^\infty \left(\frac{x}{\ell_0}\right)^{n}
\frac{\alpha_0^{n}}{\Gamma(n)} - 1\right] = \left(\frac{x}{\ell_0}\right)^{-1} \left[\exp(x\alpha_0/\ell_0) - 1\right]
\end{align}
Thus, we obtain for $p(x)$
\begin{align}
p(x) = \frac{(1 - \alpha_{0})\exp(- x(1 - \alpha_0)/\ell_0)}{\ell_0}
\left(\frac{x}{\ell_0}\right)^{-1} \left[1 - \exp(-x \alpha_0/\ell_0)\right]
+ (1 - \alpha_0) \delta(x)
\end{align}
The characteristic decay distance here is $\ell_0/(1 - \alpha)$ after which $p(x)$ decays exponentially with distance. For $\ell_0 \ll x \ll \ell_0/(1 - \alpha_0)$ it decays as $x^{-1}$. 

\subsection{Variable diving rate and Gamma-distributed pore lengths}

\begin{figure}
\includegraphics[width=0.8\textwidth]{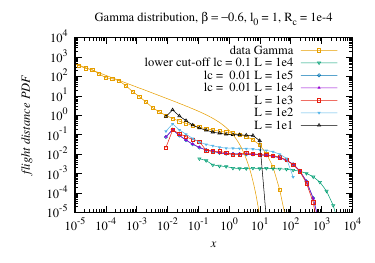}
\caption{Distribution of flight distances for a Gamma pore length distribution.\label{fig}}
\end{figure}

The probability for diving
\begin{align}
\beta(\lambda) = \frac{\pi (\lambda/2)^2 - \pi (\lambda/2-R_c)^2}{\pi (\lambda/2)^2}
H(\lambda/2 - R_c) + H(R_c - \lambda/2). 
\end{align}
The probability that the colloid does not dive is then accordingly
\begin{align}
\alpha(\lambda) = 1 - \beta = (1-2 R_c/\lambda)^2
H(\lambda/2 - R_c)  - H(R_c - \lambda/2) 
\end{align}
In this case, we obtain for the distribution of flight lengths
\begin{align}
p(x) = \left\langle \sum\limits_{n = 0}^\infty \delta\left(x - \sum\limits_{i = 1}^n \lambda_i\right) \prod\limits_{i = 1}^n (1-2 R_c/\lambda_i)^2
[1 - (1-2 R_c/\lambda_{n+1})^2] \right\rangle
\end{align}
where we assume that all pore radii are larger than the critical radius. We consider now the Laplace transform in $x$ to obtain
\begin{align}
\widetilde p(k) &= \left\langle \sum\limits_{n = 0}^\infty \exp\left(- k \sum\limits_{i = 1}^n \lambda_i\right) \prod\limits_{i = 1}^n \alpha(\lambda_i)
[1 - \alpha(\lambda_{n+1})] \right\rangle
\nonumber\\
&= \left\langle \sum\limits_{n = 0}^\infty \prod\limits_{i = 1}^n \exp\left(- k \lambda_i\right)  \alpha(\lambda_i)
[1 - \alpha(\lambda_{n+1})] \right\rangle
\nonumber\\
&= \sum\limits_{n = 0}^\infty \prod\limits_{i = 1}^n \left\langle \exp\left(- k
\lambda_i\right) \alpha(\lambda_i) \right\rangle
\left\langle [1 - \alpha(\lambda_{n+1})] \right\rangle
\nonumber\\
&= \sum\limits_{n = 0}^\infty \left\langle \exp\left(- k
\lambda\right) \alpha(\lambda) \right\rangle^n
\left\langle [1 - \alpha(\lambda)] \right\rangle
\nonumber\\
&= \frac{\left\langle [1 - \alpha(\lambda)] \right\rangle}{1 - \left\langle \exp\left(- k
\lambda\right)  \alpha(\lambda)\right\rangle}
\end{align}
We need to determine
\begin{align}
F(k)&\equiv \left\langle \exp\left(- k \lambda\right)  \alpha(\lambda)\right\rangle =
\int\limits_{0}^\infty d y p_\lambda(y) \exp\left(- k y\right)  \alpha(y)
\end{align}
in order to understand the $k$-dependence of $\widetilde p(k)$. The pore length distribution $p_\lambda(y)$ is given by the Gamma distribution,
\begin{align}
\label{eq:gamma}
p_\lambda(y) = \left(\frac{y}{\ell_0}\right)^{\beta-1}
\frac{\exp(-y/\ell_0)}{\Gamma(\beta) \ell_0}. 
\end{align}
Thus, we obtain for $F(k)$
\begin{align}
F(k) & =
\int\limits_{0}^\infty d y \left(\frac{y}{\ell_0}\right)^{\beta-1}
\frac{\exp(-y/\ell_0)}{\Gamma(\beta) \ell_0} \exp\left(- k y\right)
\alpha(y)
=
\int\limits_{0}^\infty d y \left(\frac{y}{\ell_0}\right)^{\beta-1}
\frac{\exp[-y (k + 1/\ell_0]}{\Gamma(\beta) \ell_0}  \alpha(y),
\end{align}
where $\beta = 0.4$.  This shows that $F(k) = G(k+1/\ell_0)$, that is, it is a function of $k+1/\ell_0$. This implies that $p(x)$ decays exponentially fast for $x > \ell_0$. Figure~\ref{fig} shows the distribution of flight distances for the Gamma distribution~\ref{eq:gamma} with $\beta = 0.4$ and $\ell_0 = 1$. 

\begin{figure}[htb!]
    \centering
    \includegraphics[width=0.9\linewidth]{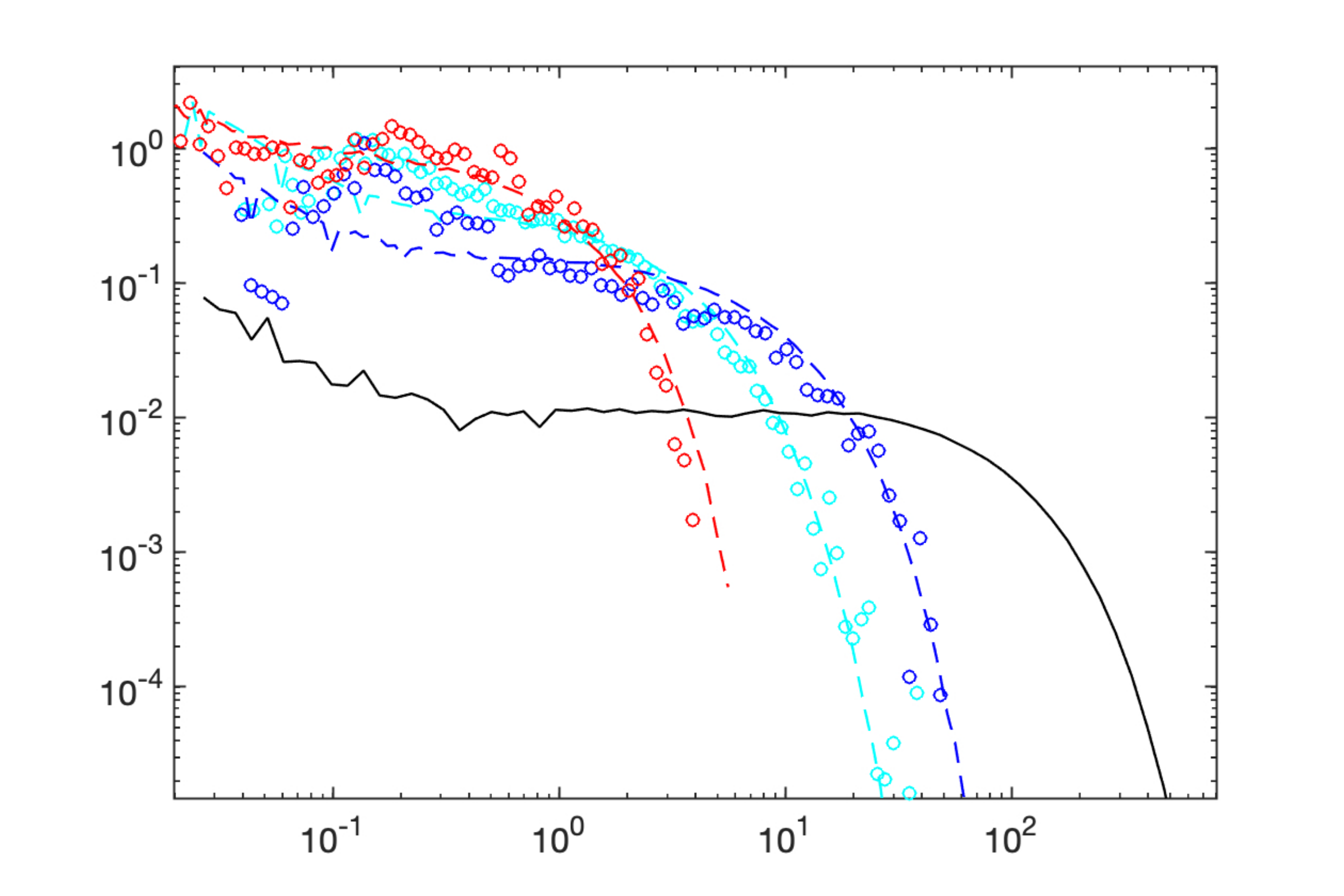}
    \caption{Measured flight size $\Lambda$ PDF. Data obtained from our particle tracking experiment are shown as red circles and from particle tracking numerical simulations (blue and cyan dots) over the velocity field in~ref{fig:simfield}. For the simulations we also computed the flight size distribution for two values of the critical distance $d_c = 0.5 \, \mu$m (blue circles) and $d_c = 1\, \mu$m (cyan circle). Each curve is normalized to unity. The black solid line represents our model prediction with $d_c = 0.01 \,\mu$m for the pore throat size distribution of our experiment: this is distribution used for the macroscopic prediction for DP and BTC.}
    \label{fig:lambda_PDF}
\end{figure}

\begin{figure}[htb!]
    \centering
    \includegraphics[width=1\linewidth]{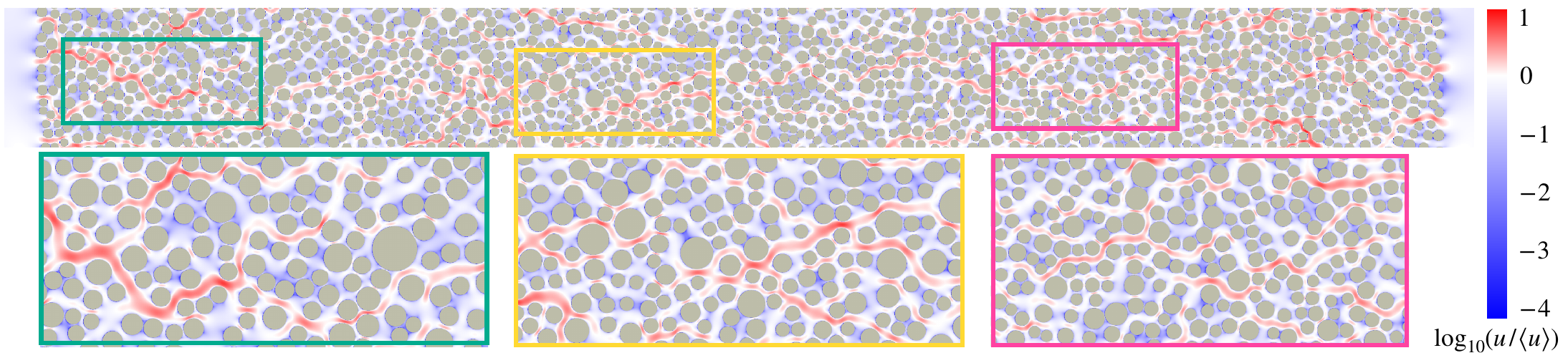}
    \caption{Modulus of the velocity field in logarithmic scale, from~\cite{deAnnaPRF2017} used for particle tracking trajectories. Particle tracking simulations have been performed to obtain $50,000$ trajectories: for each trajectory the distance from the closest grain wall have been computed and from that, the length of flights and dives, reported in figure~\ref{fig:lambda_PDF}}
    \label{fig:simfield}
\end{figure}

\subsection{$\Lambda$-PDF}

\begin{figure}
    \centering
    \includegraphics[width=0.8\linewidth]{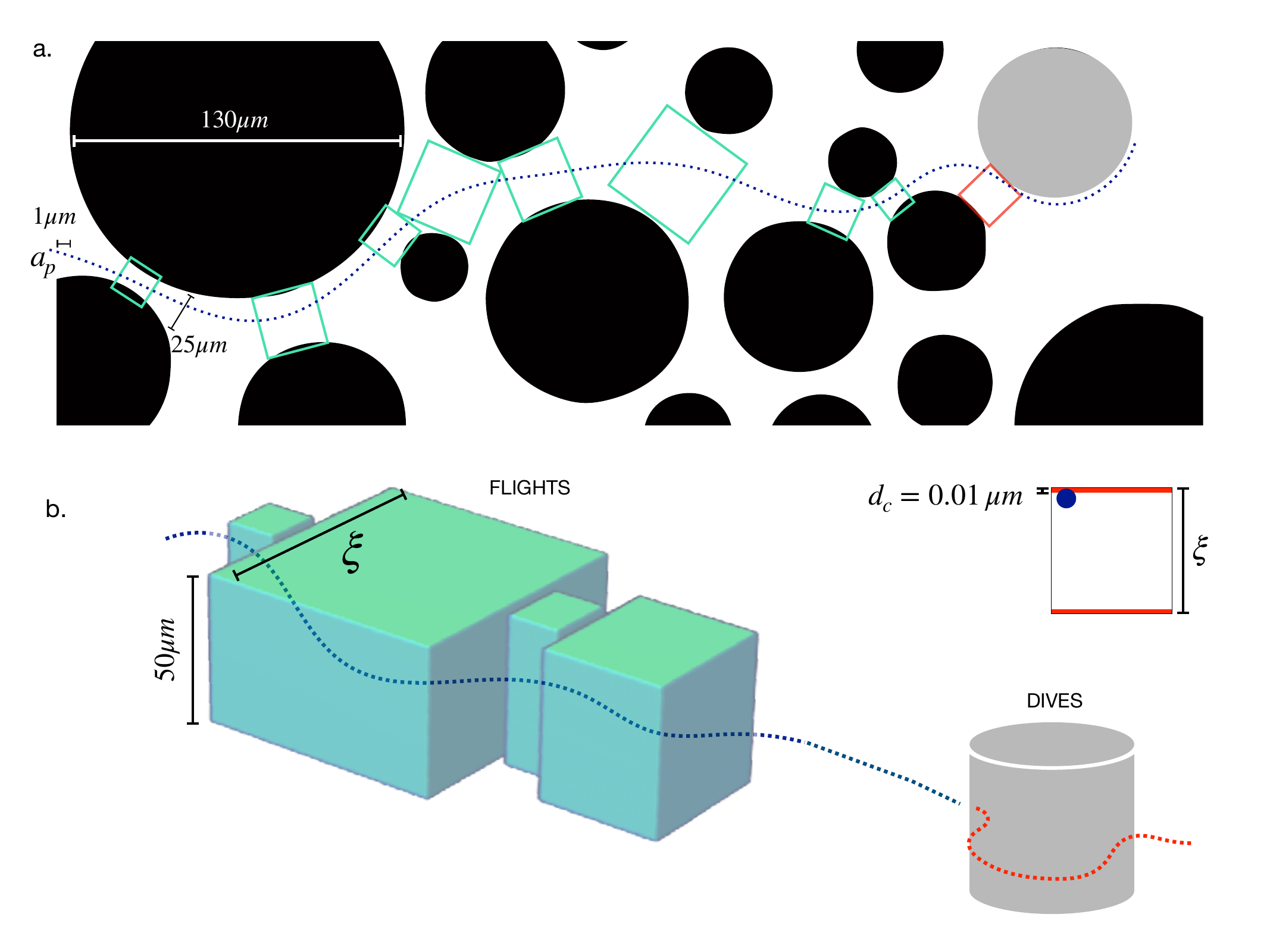}
    \caption{Schematic representation of the flow-through-pores dynamics, underpinning the stochastic model derived here to predict flights size distribution. The schematic highlight the relevant spatial scales: i) the colloid size $1 \, \mu$m, ii) microfluidics thickness, iii) pore throat size $[10-300]~\mu$m, iv) critical distance within the dive phase and v) the overall flight size (length of the dotted line).}
    \label{fig:enter-label}
\end{figure}

\begin{figure}
    \centering
    \includegraphics[width=0.8\linewidth]{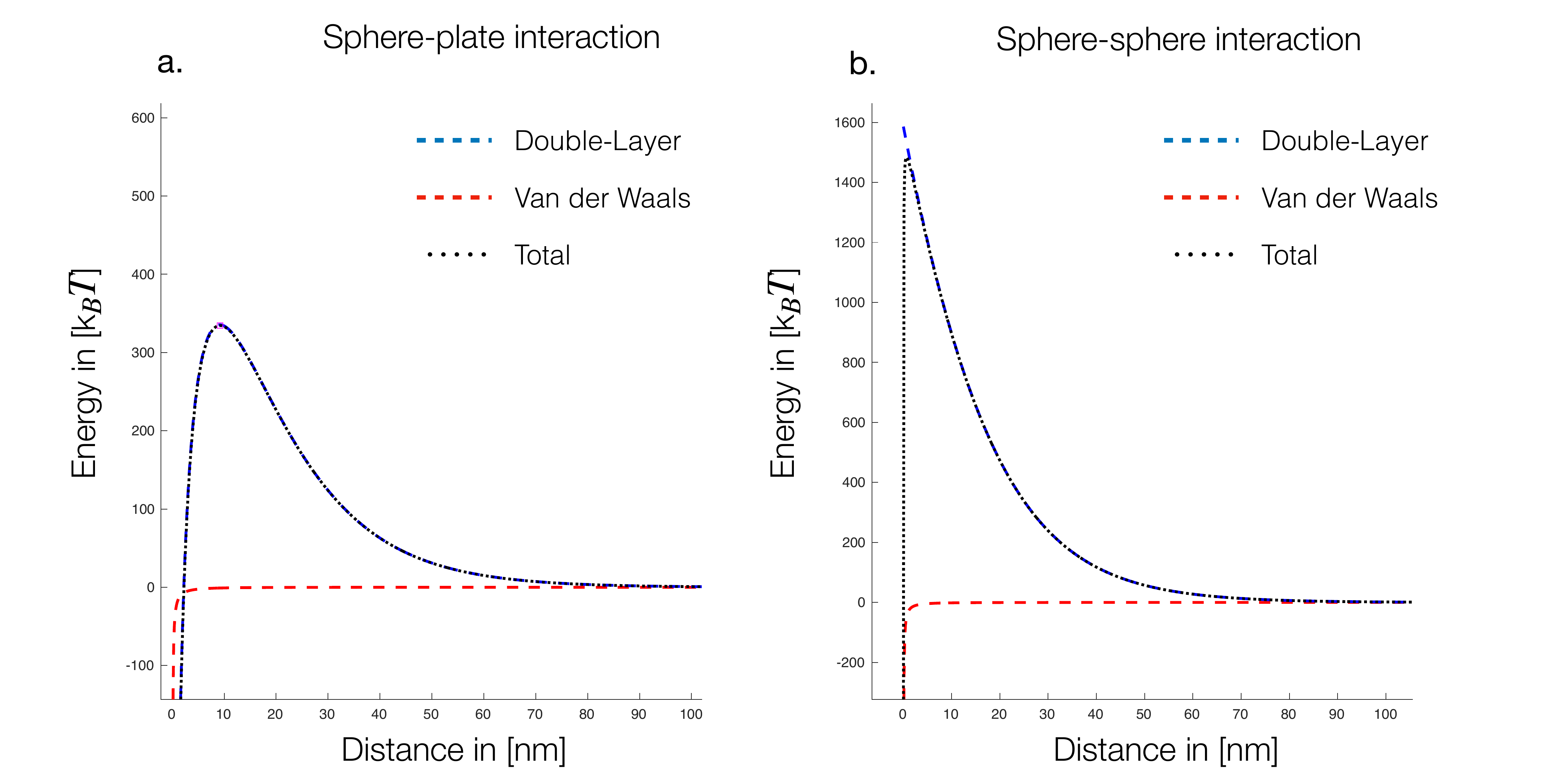}
    \caption{The total interaction energy within the particle-PDMS surface system, $U_{tot}(x)$, is the additive contribution from Van der Waals attraction \cite{Hogg1966}, $U_{VdW}(x)$ and electric double layer interaction, $U_{el}(x)$, \cite{ruckenstein_adsorption_1976}, where for Z-Potential of PDMS we set -60mV as measured by \cite{kirby_zeta_2004}, while for the Hamaker constants we set $H_c= 6.3e-20J$ and $H_{PDMS} =4.4e-20J$, \cite{yu_evaporative_2017} and $H_w = 3.7e-20J$ for latex particles, PDMS, and water respectively.}
    \label{fig:enter-label}
\end{figure}
\clearpage
\bibliography{clean_library_mf}

\end{document}